\documentclass[aps,prb,twocolumn]{revtex4}
\usepackage{amsmath}
\usepackage{natbib}
\usepackage{cases}
\usepackage[pdftex]{graphicx}

\usepackage[usenames]{color}

\newcommand{\R}[1] {{\color{black} #1}}

\begin{document}

\date{\today}

\author{ A. Shalit, \R{S. J. Mousavi}, and P. Hamm$^\ast$}

\affiliation{Department of Chemistry, University of
Zurich, Winterthurerstrasse 190, CH-8057 Zurich, Switzerland,
peter.hamm@chem.uzh.ch}

\title {2D Raman-THz Spectroscopy of Binary CHBr$_{3}$-MeOH Solvent Mixture}

\begin{abstract}
\textbf{ABSTRACT}: Hybrid two-dimensional (2D) Raman-terahertz spectroscopy is used to measure the interactions between two solvents pair in the binary CHBr$_{3}$-MeOH  mixture in the frequency range of 1-7~THz. Changes in the cross peak signature are monitored, originating from the coupling of an intramolecular  bending mode of CHBr$_{3}$  to the collective intermolecular degrees of freedom of the mixture. The appearance of a new cross peak in the 2D spectrum  measured for solvent mixture with MeOH molar fraction of 0.3 indicates a coupling to a new set of low-frequency modes formed due to  the  hydrogen bond interactions between the two solvents. This interpretation is supported by the  measurement of the CHBr$_{3}$-CS$ _{2}$ binary solvent mixture as well as by 1D absorption measurements of neat MeOH.
\end{abstract}

\maketitle

\section{Introduction}

Over the course of recent years, there has been significant progress in the development of time-resolved multidimensional spectroscopy techniques in the low-frequency part of the electromagnetic spectrum. Following the  well-established multidimensional techniques in infrared regimes,\cite{Hamm_book}  their low-frequency counterpart is aiming to access detailed structural information through  spectral densities residing below 10~THz (1~THz = 33 cm$^{-1}$), where the rotational transition of molecules in the gas phase, strong transition in the semiconductor solids, and \textit{intra}/\textit{inter}molecular vibrational motion of the liquid phase can be found.

The first attempts to study the inhomogeneous nature of the low-frequency transitions in liquids were inspired by the concept of 2D Raman spectroscopy, which was proposed  by Tanimura and Mukamel in 1993.\cite{Mukamel_1993} 2D Raman spectroscopy allows observing microscopic molecular dynamics through the change of the polarizability of the sample in response to the sequence of five electronically off-resonant excitations.\cite{kub03} Despite the tremendous experimental challenges associated with the realization of the fifth-order nonlinear technique, signals from simple organic liquids such as CS$ _{2} $ and formamide were successfully measured.\cite{kaufman02,li08} Recently, a single pulse 2D Raman technique developed by Silberberg and coworkers was implemented to overcome the most significant experimental obstacle in the form of parasitic low-order cascading process\cite{Silberberg_2015} and demonstrated the ability to measure intramolecular coupling of simple halogenated liquids.\cite{Silberberg_2020}
 Steady progress in the generation and manipulation of strong THz fields\cite{Nelson_15} stimulated the development of alternative spectroscopic path toward low-frequency multidimensional spectroscopy --  third-order 2D THz spectroscopy\cite{Elsaesser_book} which probes multipoint correlation function of the nuclear dipole moment of low-frequency transitions. 2D THz spectroscopy was successfully realized on molecules in a gas phase\cite{Fleischer_2012,Nelson_2016} and semiconductor solids\cite{Elsaesser_2011}, however, the still rather limited available energies and bandwidth of THz pulses, and the experimental limitation in generation and recombination of multiple THz pulses, have not allowed yet the realization of such experiments in liquids, where transition dipoles of the low-frequency modes are extremely small.

In an  attempt to circumvent the experimental limitation of the ``pure'' techniques mentioned above, 2D Raman and 2D THz spectroscopy,  hybrid methods, which combine indirect optical and direct THz excitation in a single experiment, were proposed and realized experimentally for  various systems, including molecular liquids\cite{Savolainen_2013,Blake_2016,Blake_2017,Shalit_2017,Grechko2018,Berger_2019,Ciardi_2019} and semiconductor solids.\cite{Johnson_2019} Bonn and coworkers combined THz, mid-IR, and visible pulses to measure couplings between the high and low-frequency modes in liquid water.\cite{Grechko2018} Blake and co-workers have introduced 2D THz Raman spectroscopy with the THz-THz-Raman (TTR) pulse sequence to study halogenated liquids.\cite{Blake_2016,Blake_2017}

At the same time in our group, an alternative hybrid method denoted 2D Raman-THz spectroscopy with the  Raman-THz-THz pulse sequence (RTT) has been under steady development.\cite{Savolainen_2013} 2D Raman-THz spectroscopy allows to interrogate the low-frequency vibrational modes of liquids consecutively through impulsive (nonlinear) Raman and direct (linear) THz excitation, providing equivalent information accessible by conventional third-order echo-based spectroscopies.\cite{Hamm_2012,Hamm_2017} By measuring the extent of the emitted echo signals, the approach demonstrated the ability to report on the inhomogeneity of low-frequency intermolecular modes of liquid water\cite{Savolainen_2013} and aqueous salt solutions, providing a glance into the correlation between macroscopic viscosity and structuring of hydrogen-bond networks on the molecular level.\cite{Shalit_2017} In  further work, we were able to disentangle the contribution of nuclear quantum effects to water's structure in different water isotopologues.\cite{Berger_2019}

We also applied 2D Raman-THz spectroscopy  to a series of halogenated organic molecules, liquid bromoform and iodoform.\cite{Ciardi_2019}
This study was inspired by the work of Blake and co-workes on similar halogenated  systems with THz-THz-Raman (TTR) pulse sequence mentioned earlier,\cite{Blake_2016,Blake_2017} to see whether RTT pulse sequence can provide complementary information on these molecules. The most recent interpretation of the TTR signals by Blake and coworkers suggested that the observed signal stems from the instantaneous excitation of intramolecular vibrations through a two-photon absorption process by a pair of strong THz fields (sum-frequency excitation pathway).\cite{Blake_2020} This explanation
is consistent with the interpretation of the THz-Kerr effect observed earlier for similar liquids.\cite{Blake_2015,Nelson_2009,Sajadi2017,Kampfrath_2018} Our RTT pulse sequence lacks the strong initial THz process required for that process. Concordantly, we suggested that the RTT response originates from a different coherence pathway, initiated by a Raman induced excitation of intramolecular modes, and followed by a weak THz interaction that switches coherences from intra to intermolecular modes through two-quantum process revealing a cross peak between these two degrees of freedom.\cite{Ciardi_2019}

In this paper, we aim to test this interpretation further by measuring 2D Raman-THz spectra for a binary bromoform-methanol (CHBr$_{3}$-MeOH) solvent mixture, which is more complex with respect to its intermolecular response. To that end, we monitor changes in the cross peaks signature originating from the coupling of intramolecular symmetric bending mode of CHBr$_{3}$ at \textit{v}$_{3}$ = 6.7~THz with the broad intermolecular band at about 1.5~THz as a function of the MeOH concentration in the CHBr$_{3}$-MeOH mixture.

It is well known that various excess thermodynamic properties such as the molar heat of mixing (\textit{H}$^{E}$), molar excess Gibbs free energy (\textit{G}$^{E}$), and excess volumes of mixing (\textit{V}$^{E}$) for binary solvent systems of $n$-alcohols with CHCl$_{3}$ and CHBr$_{3}$ show significant deviation from ideal behavior.\cite{Sidhu_1978} For example, \textit{V}$^{E}$ for the CHBr$_{3}$-MeOH solvent mixture is constantly negative throughout the entire mole fraction range,\cite{Sidhu_1978_2} indicating that there are specific interactions between alcohol chains and CHBr$_{3}$, which result in the volume decrease (negative \textit{V}$^{E}$). They overweight positive \textit{V}$^{E}$ factors such as the steric repulsion between the alcohol alkyl chains and the bulky Br atom, and the rupture of the MeOH hydrogen bond networks upon dilution with CHBr$ _{3}$.  It was proposed that the hydrogen bond donating ability of CHBr$ _{3}$ (the H atom) and the hydrogen bond accepting  properties of MeOH  (the O atom) are primarily responsible for the formation of networks of hydrogen bond clusters in CHBr$_{3}$-MeOH binary system. Further UV-Vis studies with various solvatochromic probe molecules in CHCl$_{3}$-MeOH mixture,\cite{Gupta_2012} and its dynamical characterization by means of femtosecond transient absorption spectroscopy,\cite{Gupta_2015} provided clear evidence that both molecules interact through hydrogen bond networks. \R{Molecular dynamic simulations studies of the binary mixtures  with unusual mixing behavior such as  CHCl$_{3}$-MeOH and DMSO-water were carried out to quantify hydrogen-bond populations, providing insight into co-solvents  interactions on the  molecular level.\cite{Gratias_1998,Oh_2017}}  In the current study, we measure that interaction via a cross peak between a intramolecular mode of CHBr$_{3}$ and the intermolecular networking modes of the binary mixture directly in the low-frequency THz spectral range.

\section{ Methods}

The experimental set-up for 2D Raman–THz spectroscopy was essentially the same as described in detail before,\cite{Savolainen_2013} albeit based on a different laser system. In brief, the output of a Yb-doped fiber laser/amplifier system (short-pulse Tangerine, Amplitude Systems, France) with a central wavelength of 1030~nm,  pulse duration of 150~fs, and repetition rate of 10~KHz was split into THz and Raman branches. The THz pulses were generated via optical rectification by focusing 7~$\mu$J of the fundamental amplifier output into a 100~$\mu$m thick (110) GaP crystal, which generated weak but broadband THz pulses with a near single-cycle shape and a bandwidth that extends to $\sim$7~THz \R {(see Figure S1}). The THz pulse was focused on the sample by a custom made elliptical mirror with 2$f$ = 83~mm  and the emitted signal was subsequently detected with an equivalent GaP crystal via electro-optic sampling with enhanced sensitivity.\cite{Ahmed_2014}
Raman pulses  with a central wavelength of 860~nm,  \R {bandwidth of $\sim$ 9~THz}, and energy of 5~$\mu$J were produced in an OPA (Twin STARZZ, Fastlight, France) pumped from the second harmonic of the Tangerine system.

Delay \textit{t}$_{1}$ between the Raman pump and the THz pulse was controlled by a step-scan motor (Physics Instruments, M-405.DG), while sampling delay \textit{t}$_{2}$  was scanned contentiously by a fast-scanning motor (Physics Instruments, V-408). The sample was contained in the static cuvette made of two thin sapphire windows (UQG optics) separated by a 300~$\mu$m Teflon spacer.  \R{ Data were zero-padded in both temporal dimensions, and a 2D low-pass filter with a cutoff frequency of 7 THz (determined by the available bandwidth of our spectrometer) was applied to all data. All data presented in the manuscript were collected with the same temporal window and Fourier transformed consistently.}

The averaging times varied from 5~h for the pure CHBr$_{3} $ sample  \R{up to a total of  80~h for the X$_{MeOH}$ = 0.3 sample. The latter was measured independently three times (see Figure S4 for the individual X$_{MeOH}$ = 0.3 measurements, evidencing the reproducibility of these results). An average over those three measurements will be shown here.} \R{Transmission of the THz field through the sample was constantly monitored  during the signal acquisition to verify the molar content of the MeOH}.

In contrast to our previous work,\cite{Ciardi_2019} we undersampled data along the Raman axis by a factor of 2. This reduced the signal acquisition times by the same factor, which became crucial when the much smaller signals from CHBr$ _{3}$-MeOH mixtures are measured. There are two intramolecular vibrational modes (asymmetric \textit{v}$_{6}$ = 4.7~THz and symmetric \textit{v}$_{3}$ = 6.7~THz C-Br bending modes) giving rise to two cross peaks in the 2D Raman-THz spectrum due to the coupling to the intermolecular modes of CHBr$_{3}$ at $\sim$ 1.5~THz .\cite{Ciardi_2019} The \textit{v}$_{3} $ mode tends to provide much stronger cross peaks signals, despite the fact that it is placed on the edge of the detection bandwidth of our 2D spectrometer.  Under-sampling along the Raman axis results in the ``folding'' of the 6.7~THz mode into the rephasing part of the 2D spectra and inherent separation of that signal from the weak contribution of the \textit{v}$ _{6}$ mode. \R{Such an under-sampling acquisition is allowed due to inherent suppression of the rephrasing part of the emitted signal by the specific spectral shape of the  instrument response function (IRF) of our 2D Raman–THz spectrometer (for a detailed discussion see.\cite{Ciardi_2019})}. In all representations of 2D spectra, the data are folded back into the non-rephasing quadrant.

\section{Results and Discussion}

\begin{figure}[t]
	\centering
	\begin{center}
		\includegraphics[width=.5\textwidth]{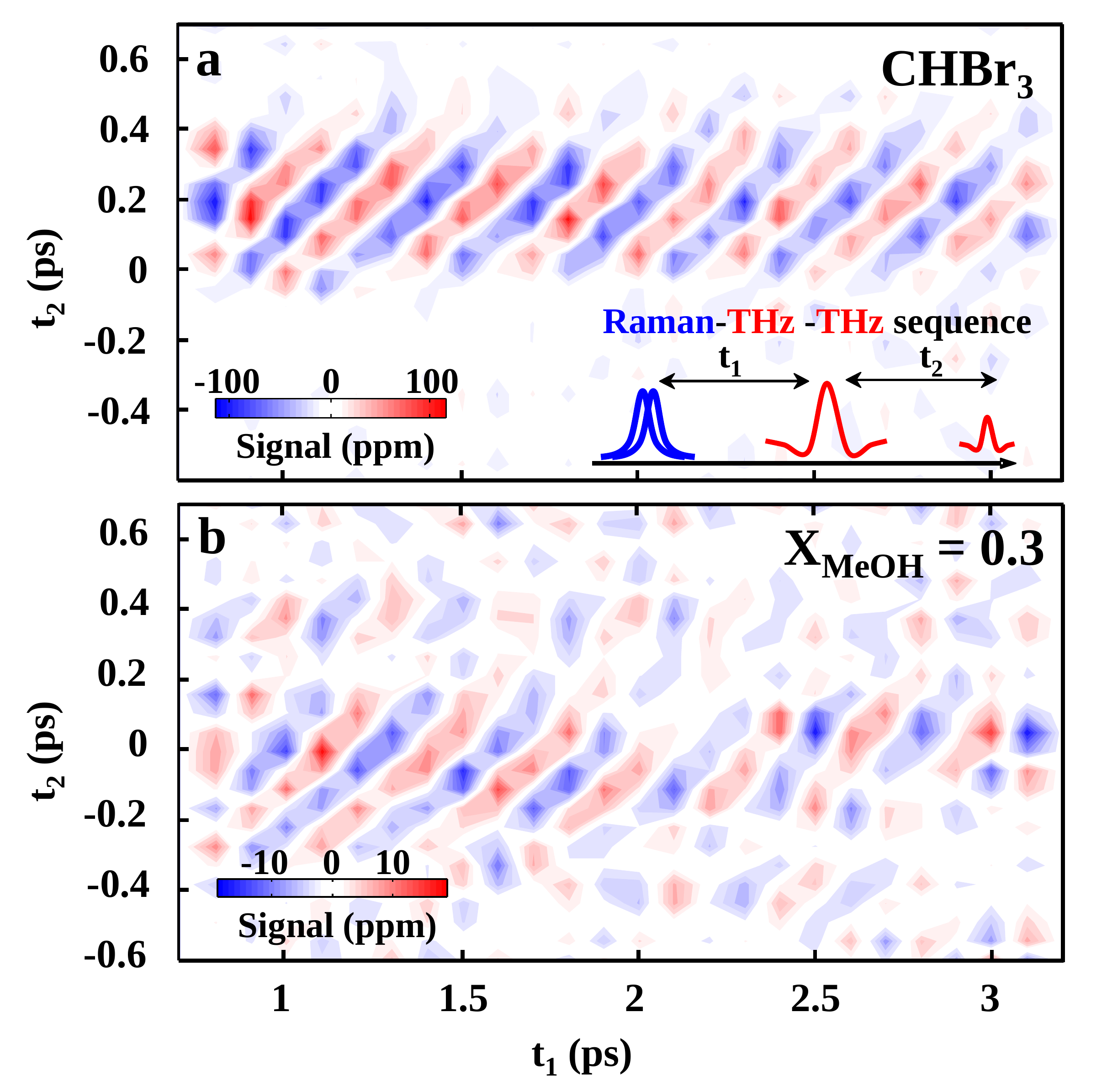}
		\caption{Experimental 2D Raman-THz data for a) neat CHBr$_{3}$ and b) CHBr$_{3}$-MeOH binary mixture with X$_{MeOH}$ = 0.3.}\label{Fig1}
	\end{center}
\end{figure}

\begin{figure}[t]
	
	\centering
	
	\begin{center}
		
		\includegraphics[width=.45\textwidth]{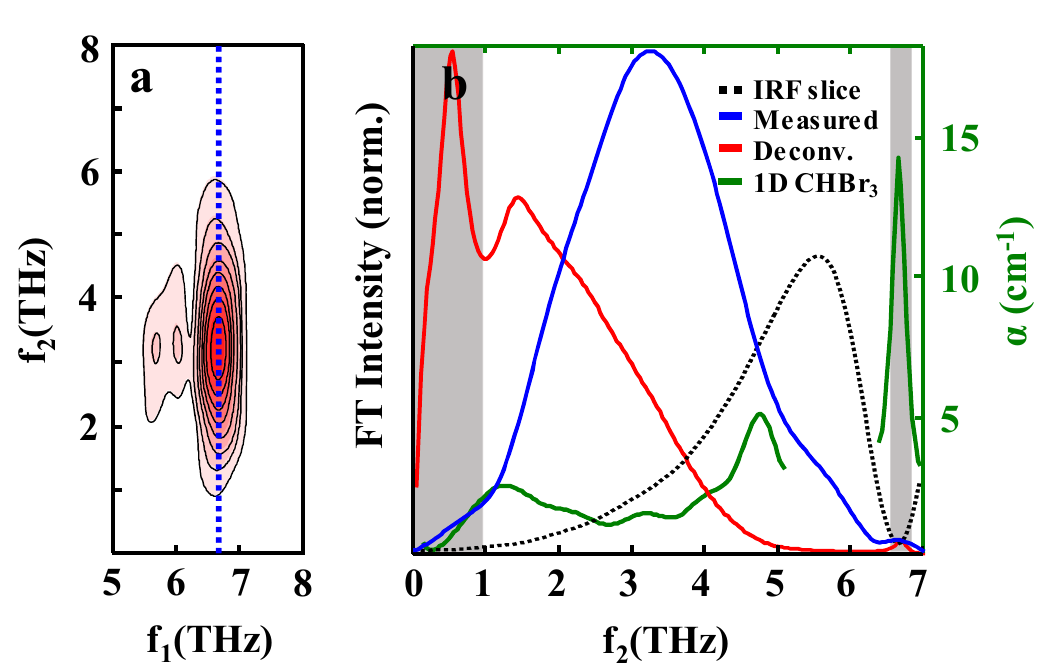}
		
		\caption{a) Absolute value of the 2D Fourier transform for the neat CHBr$_{3}$ signal in Figure~\ref{Fig1}a. (b) 1D vertical cut through the 2D spectrum at \textit{f}$_1$ = 6.7~THz (blue line), together with the deconvoluted data (red line) obtained by dividing through the IRF slice (black dashed line),  and linear THz spectrum of neat CHBr$_{3}$ (green line ). The regions with unreliable deconvolution are shaded out  \R{ and  region contaminated by the response of the sapphire window was removed from the 1D THz absorption spectrum of CHBr$_{3}$}.}\label{Fig2}
		
	\end{center}
	
\end{figure}

We start in Figures~\ref{Fig1}a and \ref{Fig2} with neat CHBr$_{3}$, which we have also investigated in our previous works,\cite{Ciardi_2019} but has been measured again for a consistent comparison with the subsequent dilution series. Figure~\ref{Fig1}a shows its 2D Raman-THz signal in the Raman-THz-THz quadrant measured beyond the pulse overlap region (\textit{t}$ _{1} $$>$ 0.7~ps). The vibrational signal was isolated from rotational contributions, which originate from the alignment of the CHBr$_{3}$ molecules along the polarization direction of the strong Raman pulse, by substracting a single-exponential fit along each \textit{t}$_{1}$ cut. There is a very long-lived  oscillatory signal along the \textit{t}$ _{1} $  direction (Raman to THz delay), extending significantly beyond our measurement window, while the signal decays significantly faster along \textit{t}$_{2}$ axis (THz to THz delay). It appears to be a rephasing signal with the fringes inclined along the diagonal, but that is an artifact from the undersampled acquisition along \textit{t}$_{1}$ (see \textbf{Methods}). Figure~\ref{Fig1}b depicts the signal obtained for the CHBr$_{3}$-MeOH binary solvent mixture with the highest MeOH molar fraction X$_{MeOH}$ = 0.3 used At the same timework. Two main differences can be noticed upon MeOH dilution: a significant decrease in signal size (more than a factor of 5) and additional beating of the oscillatory signal along the THz (\textit{t}$ _{2}$) axis with a nodal line around \textit{t}$_{2}$ = 0.25~ps.  Both effects will be quantified and discussed in detail below.

Figure~\ref{Fig2}a shows the absolute value of the 2D Fourier transformation of the data in Figure~\ref{Fig1}a, revealing a single cross peak with the \textit{f}$_{1}$ frequency matching exactly the \textit{v}$_{3}$ symmetric bending mode of CHBr$_{3}$. Along the \textit{f}$_{2}$ axis, it shows a broad band peaking at around 3.3~THz, as is evident from the cut in Figure~\ref{Fig2}b, blue line. As  discussed in great details in our previous publication,\cite{Ciardi_2019} the convolution of the molecular response with the IRF significantly distorts position, shape, and amplitude of the observed peaks in the 2D spectrum. In order to decipher the real frequency  position of the observed cross peaks, one needs to reconstruct the IRF by taking into consideration temporal shapes of THz and Raman pulses, and subsequently perform a deconvolution by dividing the measured response through IRF in the frequency domain. \R {In Figure ~\ref{Fig2}b, the red line shows the result of that procedure, with the cross peak shifted downwards significantly along the \textit{f}$_{2}$ axis, now peaking at 1.5 THz. This coincides with vibrational modes, which are commonly assigned to the collective intermolecular motions of CHBr$_{3}$, and which can be seen in 1D THz abortion spectrum (Figure ~\ref{Fig2}b, green line) and many other polar liquids.\cite{Davies_1968,Afsar_1975}}

\begin{figure}[t]
	
	\centering
	
	\begin{center}
		
		\includegraphics[width=.40\textwidth]{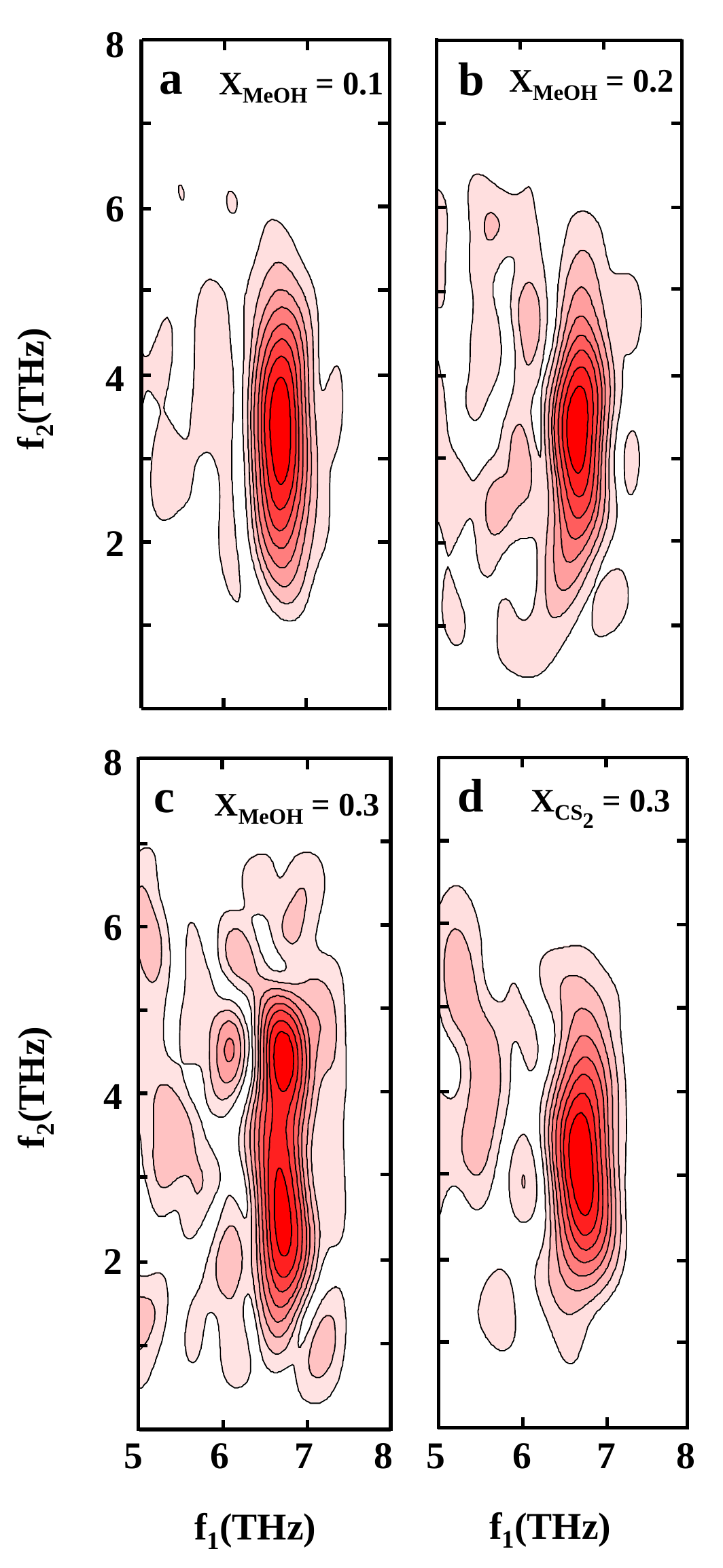}
		
		\caption{Absolute value 2D spectra of a series of  CHBr$_{3}$-MeOH binary mixtures with a) X$_{MeOH}$ = 0.1, b) X$_{MeOH}$ = 0.2, and c) X$_{MeOH}$ = 0.3 molar fractions. (d) Absolute value 2D spectrum of the control  CHBr$_{3}$-CS$_2$ binary mixture with X$_{CS_{2}}$ = 0.3.}  \label{Fig3}
		
	\end{center}
	
\end{figure}

Figures~\ref{Fig3}a-c show the absolute value 2D spectra obtained for a series of CHBr$_{3}$-MeOH mixtures with X$_{MeOH}$ = 0.1, 0.2, and 0.3 mole fraction of MeOH. Figure~\ref{Fig3}d depicts the 2D spectrum obtained for a solution with 0.3 mole fraction of CS$_{2} $ in CHBr$_{3}$, which was measured as a control.  Comparison with the reference data of neat CHBr$_{3}$ (Figure~\ref{Fig2}a) shows that the addition of  MeOH with 0.1 and 0.2 mole fraction (Figures~\ref{Fig2}a and b) does not significantly affect the shape of the cross peak. The only significant difference is the reduction in signal size upon addition of MeOH.  However, for the X$_{MeOH}$ = 0.3 (Figure~\ref{Fig3}c), along with a further reduction of signal size, there is a significant change in the cross peak line-shape, which splits into two portions and becomes significantly broader. \R{The origin of this split is the result of the coherent beating discussed in the context of the time-domain data of Figure~\ref{Fig1}b.} On the other hand, the control measurement of a binary mixture with 0.3 mole fraction of CS$_{2}$ does not show any significant change in the cross peak line-shape. \R{It is important to note that we observe a quite non-trivial concentration dependence in the 2D response, while the linear absorption spectra (Figure~S4) seem to be simply additive.}

\begin{figure}[t]
	
	\centering
	
	\begin{center}
		
		\includegraphics[width=.4\textwidth]{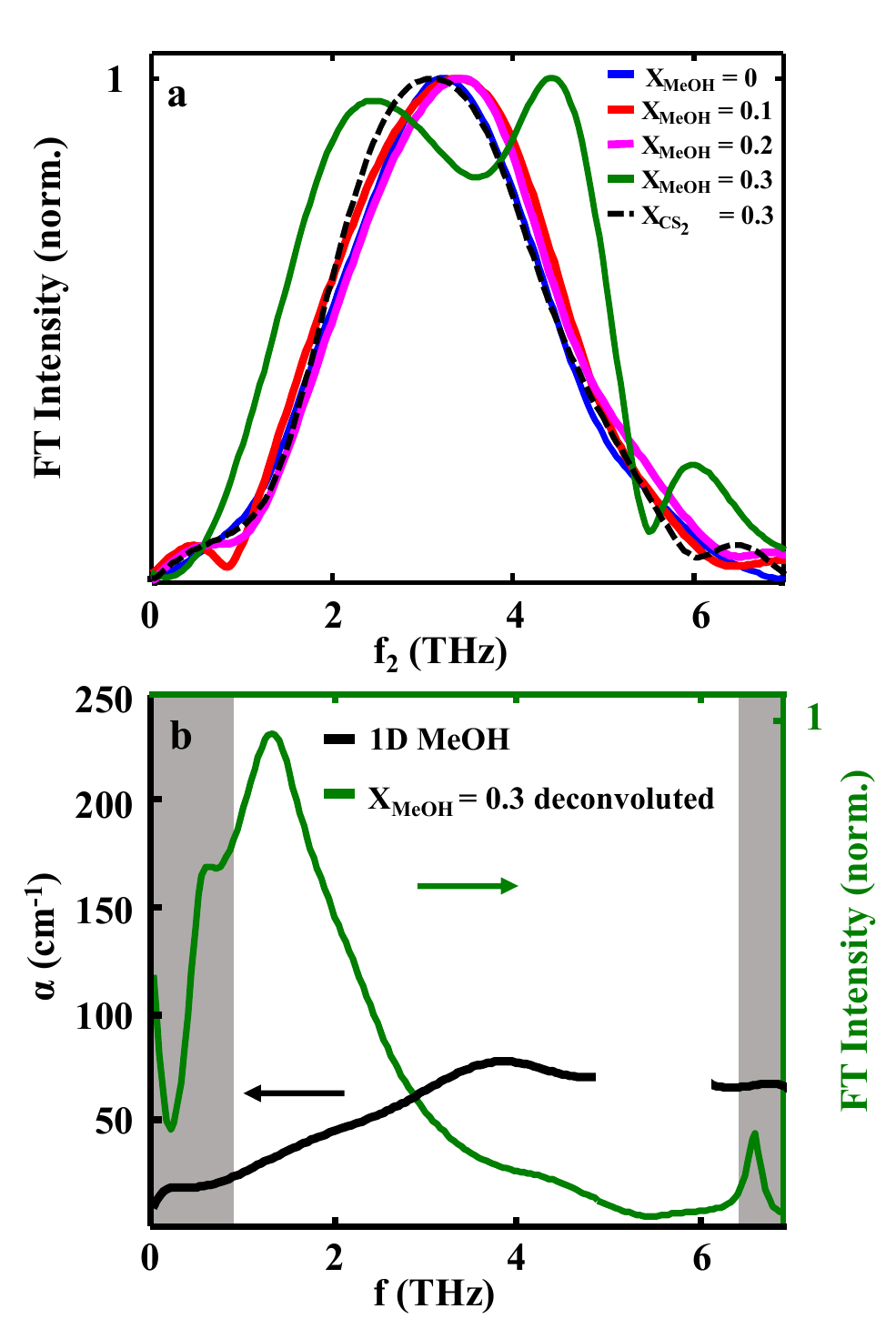}
		
		\caption{a) Comparison of  vertical cuts corresponding to \textit{f}$_1$ = 6.7~THz for neat CHBr$ _{3} $ (blue), X$_{MeOH}$ = 0.1 (red), X$_{MeOH}$ = 0.2 (magenta), X$_{MeOH}$ = 0.3 (green), and X$_{CS_2}$ = 0.3 (black dashed). b) Comparison between absorption spectrum of MeOH (black line) and deconvoluted vertical cut of  X$_{MeOH}$ = 0.3 measurement at \textit{f}$_1$ = 6.7~THz. The regions with unreliable deconvolution are shaded out.  \R {The response of the sapphire windows of the cuvette was removed form 1D absorption spectrum of MeOH}.} \label{Fig4}
		
	\end{center}
	
\end{figure}

Figure~\ref{Fig4}a confirms that observation on a more quantitative level, depicting spectral cuts at \textit{f}$ _{1}$ = 6.7~THz along the \textit{f}$ _{2} $ axis for all five aforementioned measurements. The cut for the X$_{MeOH}$ = 0.3 measurement  (green line) shows two peaks around 2.4~THz and 4.4~THz,  while the cross peak remains  in the 3.1--3.4~THz region for the rest of the measurements, including X$_{CS_{2}}$ = 0.3. As was mentioned earlier, the positions of the peaks in the 2D Raman-THz spectrum are significantly distorted and deconvolution is required to obtain the real molecular response. To get a qualitative idea about the possible origin of the splitting observed in the X$_{MeOH}$ = 0.3 sample, it is useful to analyze the spectral  shape of the cross peak after deconvolution, i.e., after division by the corresponding cut of the IRF (Figure~\ref{Fig2}b, dashed black line).  The result is shown in Figure~\ref{Fig4}b (green line), whose main peak is narrower and at lower frequency (1.3~THz vs 1.5~THz) as compared to neat CHBr$_{3}$ (Figure~\ref{Fig2}a, red line). In addition, a small shoulder shows up around 4~THz. In order to understand the molecular origin of this shoulder, we performed a linear THz absorption measurement of a neat MeOH sample, which is depicted as  black line in Figure~\ref{Fig4}b. The THz spectrum of MeOH reveals a broad peak around 3.9~THz,  which was reported in the literature before\cite{SARKAR_2017} and is usually  assigned, at least partially, to  the collective hydrogen-bond stretching modes in MeOH.\cite{Vallauri_2001,Woods_2005} \R{ The de-convoluted responses of the lower MeOH fractions and control measurements are depicted in Figure S3. While small variation in the width and position of the main peak around 1.5~THz  might be interesting to investigate further, it is beyond the experimental capabilities of our 2D Raman THz spectrometer in its current state.}    

\begin{figure}[t]
	
	\centering
	
	\begin{center}
		
		\includegraphics[width=.5\textwidth]{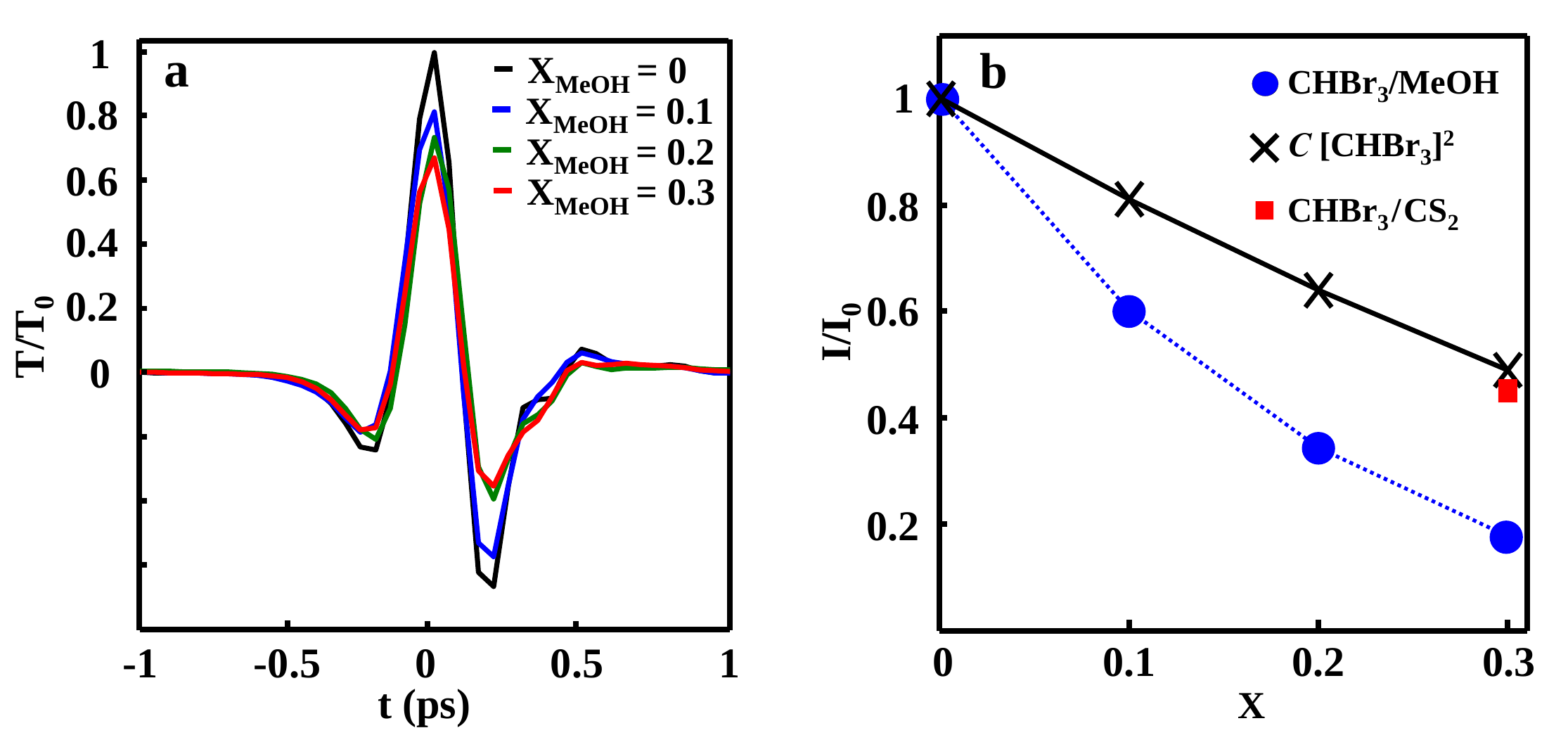}
		
		\caption{a) \R{Linear transmission of the THz field through the samples with  X$_{MeOH}$ = 0.1 (blue),   X$_{MeOH}$ = 0.2 (green), and  X$_{MeOH} $= 0.3 (red) normalized to the transmittance of the pure CHBr$ _{3} $ sample (black).}  b) Non-linear 2D Raman-THz  size dependence on CHBr$ _{3} $ dilution. Measured signal vs. MeOH mole fraction (blue circles).  Expected square concentration dependence (black crosses). Measured signal size for the control measurement with X$_{CS_{2}}$ = 0.3 (red square).} \label{Fig5}
		
	\end{center}
	
\end{figure}

The fact that the frequency  of the  shoulder in the deconvoluted 2D Raman-THz response falls very close to the intermolecular vibration of MeOH provides evidence for the interaction between the two solvents, indicating that the intramolecular bending mode of CHBr$_{3}$ is weakly coupled to the intermolecuar hydrogen bond motion of MeOH. This claim can be supported further by considering the signal obtained from the control CHBr$_{3} $--CS$_2$ binary mixture (Figure~\ref{Fig3}d). Similarly to the CHBr$_{3}$-MeOH mixture, CHBr$_{3}$ and CS$_{2}$ are fully miscible for all mole fractions. However, unlike MeOH, CS$_{2}$ lacks the ability  to form any hydrogen bond with CHBr$_{3}$ and thus can be considered a simple dilution agent. Figure~\ref{Fig3}d shows that the line-shape of the cross peak remains intact upon the addition of CS$ _{2}$, indicating that splitting observed in Figure~\ref{Fig3}c for X$_{MeOH}$ = 0.3 mixture is not related to CHBr$_{3}$ dilution but originates from the  interactions between the two solvents.

Additional indirect evidence for non-ideal behaviour is derived when examining the dependence of the signal size of the 2D response as a function of concentration of MeOH, which is significant. By the time the molar content of MeOH reaches 0.3, the non-linear 2D Raman-THz signal is decreased by a factor 5, significantly limiting our ability to measure  CHBr$_{3}$-MeOH systems beyond this composition. \R{At the same time, the linear transmittance of the THz field through X$_{MeOH}$ = 0.3 binary mixture is reduced only by $\sim$ 30\%,  as shown in Figure~\ref{Fig5}a, confirming that this sample can be still considered  optically thin and no significant effects of the re-absorption are expected. This behavior is quantified  in Figure~\ref{Fig5}b, where the signal size, scaled with the transmitted THz field (Figure~\ref{Fig5}a) and normalized to the maximum signal obtained for the neat CHBr$_{3}$ sample (Figure~\ref{Fig1}a)}, is plotted against the MeOH fraction. The signal decreases much faster than the square of concentration (black line), which would be expected in the case of a simple dilution. \R{Such a quadratic concentration dependence is expected, because cross peak signal originates from the interaction of bath modes of the solvent with an intramolecular mode of bromoform. Dilution will lead to a decrease of the concentration of the  CHBr$_{3}$ molecules, along with a disruption  of long-range interaction of the solvent, leading to a bimolecular concentration dependence.} It is worth noting that when MeOH is replaced with CS$_{2}$ (red square), this trivial concentration dependence is indeed observed. We believe that the ``missing'' signal in the X$_{MeOH}$ = 0.3 sample is an indication of the significant disruption of the long-range interaction in CHBr$_{3}$ on account of the formation of new intermolecular interaction in the CHBr$_{3}$-MeOH binary solvent.

\section{Conclusion}

The combination of experimental findings presented here, namely, the split of the \textit{v}$_{3} $ cross peak into two contributions in the X$_{MeOH}$ = 0.3 sample,  the close correspondence of one of these cross peaks with the linear MeOH spectrum, and the non-ideal signal size dependence on MeOH concentration, provide new insights into the structure and dynamics of binary solvent mixtures. \R{A deeper understanding of these effects will require molecular dynamics simulations, in connection with procedures to calculate the 2D-Raman THz response.\cite{Hasegawa2006,ito14,Hamm_2014,Blake_2019} Such MD simulations will be demanding, since they will have to build on a polarizable and flexible force field, and since convergence of the required three-point correlation functions scales very unfavorably with the coherence time of the system (which is long in the case of the intermolecular mode of CHBr$_{3}$). We nevertheless hope that our work triggers effort in that direction.}
In any case, the double peak structure of the cross peak, or correspondingly, the beating of the time-domain data in the $t_2$-direction, emphasize that the response is not instantaneous in the THz dimension in our RTT pulse sequence, unlike for the TTR pulse sequence.\cite{Blake_2020}
The demonstrated sensitivity of 2D Raman-THz spectroscopy toward low-frequency intra/intermolecular vibrational couplings can be utilized further to study structural characteristics of hydrogen bond network forming systems.

\vspace{1cm}\noindent\textbf{Supporting Information:} Figure S1: Temporal  and spectral profiles of the THz field generated by  100~$\mu$m thick (110) GaP crystal. Figure S2: Absolute value 2D Raman-THz spectra obtained from three independent, non-consecutive measurements of X$_{MeOH}$ = 0.3 sample. Figure S3: Deconvoluted spectral lineshapes of the cross-peak at  \textit{f}$_1$ = 6.7~THz  for   X$_{MeOH}$ = 0.1, 0.2,  and  X$_{CS_2}$  = 0.3 mixtures. Figure S4: 1D absorption spectrum of the binary  CHBr$_{3}$-MeOH mixtures.

\vspace{1cm}\noindent\textbf{Acknowledgement:} The work has been supported by the Swiss National Science Foundation (SNF) through the National Center of Competence and Research (NCCR) MUST as well by the MaxWater network of the Max Planck Society.\\

\makeatletter
\def\@biblabel#1{(#1)}
\makeatother

\def\bibsection{\section*{}} 

\noindent\textbf{References:}
\vspace{-1.5cm}


\providecommand{\latin}[1]{#1}
\makeatletter
\providecommand{\doi}
  {\begingroup\let\do\@makeother\dospecials
  \catcode`\{=1 \catcode`\}=2 \doi@aux}
\providecommand{\doi@aux}[1]{\endgroup\texttt{#1}}
\makeatother
\providecommand*\mcitethebibliography{\thebibliography}
\csname @ifundefined\endcsname{endmcitethebibliography}
  {\let\endmcitethebibliography\endthebibliography}{}

\newpage
\clearpage

\textbf{TOC Graphic:}

	\centering
	
	\begin{center}		
		\includegraphics[width=.5\textwidth]{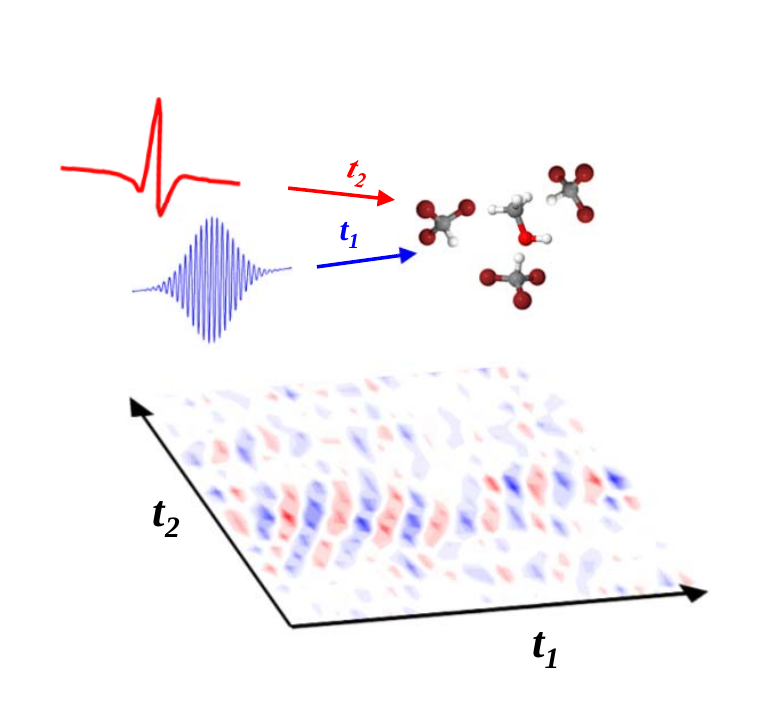}		
		
	\end{center}

\end{document}